\documentclass{jpp}
\usepackage{graphicx}

\usepackage[utf8]{inputenc}
\usepackage[T1]{fontenc}
\usepackage{amsmath}
\usepackage{multicol}
\usepackage{cleveref}
\usepackage{float}
\usepackage{algorithm}
\usepackage{algpseudocode}

\shorttitle{Data Driven Multi-fluid 10-Moment Closure}
\shortauthor{J. Donaghy, K. Germaschewski}

\title{In Search of a Data Driven Symbolic Multi-Fluid 10-Moment Model Closure}

\author{John Donaghy\aff{1}
  \corresp{\email{john.donaghy@unh.edu}},
  Kai Germaschewski\aff{1}
 }

\affiliation{\aff{1}Space Science Center, University of New Hampshire, Durham, NH 03824, US}

\renewcommand{\Vec}{\mathbf}
\DeclareMathOperator{\sech}{sech}

\begin{document}

\maketitle

\begin{abstract}
The inclusion of kinetic effects into fluid models has been a long standing problem in magnetic reconnection and plasma physics.  Generally the pressure tensor is reduced to a scalar which is an approximation used to aid in the modeling of large scale global systems such as the Earth's magnetosphere.  This unfortunately omits important kinetic physics which have been shown to play a crucial role in collisionless regimes.  The multi-fluid 10-moment model on the other-hand retains the full symmetric pressure tensor.  The 10-moment model is constructed by taking moments of the Vlasov equation up to second order, and includes the scalar density, the vector bulk-flow, and the symmetric pressure tensor for a total of 10 separate components.  Use of the multi-fluid 10-moment model requires a closure which truncates the cascading system of equations.  Here we look to leverage data-driven methodologies to seek a closure which may improve physical fidelity of the 10-moment multi-fluid model in collisionless regimes. Specifically we use the Sparse Identification of Nonlinear Dynamics (SINDy) method for symbolic equation discovery to seek the truncating closure from fully kinetic particle-in-cell simulation data, which inherently retains the relevant kinetic physics.  We verify our method by reproducing the 10-moment model from the PIC particle data and use the method to generate a closure truncating the 10-moment model which is analyzed through the nonlinear phase of reconnection. 
\end{abstract}

\section{Introduction}\label{sec:introduction}
%Reference sections like this: see \S\ref{sec:experiments} 

Magnetic reconnection is the process in which a magnetic field embedded in plasma undergoes a topological restructuring. This process, which converts the energy stored in magnetic fields into particle kinetic and thermal energy is ubiquitous throughout the universe and plays an important role in such diverse events as sawtooth crashes in fusion plasmas~\citep{Yamada2010,Zweibel2009}, magnetic substorms in the earth's magnetosphere~\citep{Hastie1997}, and solar coronal mass ejections~\citep{Masuda1994}.

Magnetic reconnection was first studied using magnetohydrodynamics (MHD). In ideal MHD, which describes a plasma as a single fluid of infinite conductivity,  Ohm's Law in the lab frame is given by $\Vec{E} + \Vec{u}\times\Vec{B} = \eta\Vec{J} = 0$ and implies that magnetic flux is frozen into the plasma flows.  Hence magnetic reconnection is topologically prohibited. 

In real plasmas, however, the right-hand-side (r.h.s) of the Ohm's Law is not zero. In resistive MHD, finite conductivity is introduced as the resistive term $\eta\Vec{J}$. Sweet and Parker~\citep{Parker1957} used this model to develop  the first self-consistent description of magnetic reconnection. However, many plasmas of interest are nearly collisionless and the reconnection rates predicted by Sweet-Parker do not match observations. Petschek's model~\citep{Petschek1964} addressed the shortcomings of the Sweet-Parker model by realizing that the geometry of the reconnection region is crucial and that reducing the aspect ratio of that region can explain the observed faster reconnection rates. This means that the reconnection region does not span the global length scale but is much shorter and connected to the global geometry by slow shocks. Petschek's model, however is not realized in numerical simulations unless one triggers a shorter reconnection region by anomalous resistivity or other modifications of the physics.

It was later realized that the one-fluid MHD description of plasmas is too limited to incorporate all the essential reconnection physics. The next steps were Hall-MHD and extended MHD models that take into account that electrons and ions do not move together at the small scales of a reconnection region. From the electron momentum equation, one can derive a generalized Ohm's Law~\citep{Vasyliunas1975,Birn2001}:

\begin{equation}
    \Vec{E} + \Vec{u}\times\Vec{B} = \eta\Vec{J} + \frac{\Vec{J}\times\Vec{B}}{n|e|} - \frac{\nabla\cdot \Vec{P}_e}{n|e|} + \frac{m_e}{n|e|^2} \times \left[\partial_t \Vec{J} + \nabla \cdot \left(\Vec{u}\Vec{J} + \Vec{J}\Vec{u} - \frac{\Vec{J}\Vec{J}}{n|e|} \right) \right]
    \label{eqn:generalized_ohms}
\end{equation}
The generalized Ohm's law~\eqref{eqn:generalized_ohms} describes the various non-ideal contributions to the electric field in a plasma.  The resistive dissipation term, sometimes denoted as the collisional term, is given by $\eta\Vec{J}$.  The second term on the right hand side is called the Hall term which in and of itself cannot support magnetic reconnection.  It is however known to change the geometry of the reconnection process to enable "fast" reconnection, where the time scales of the process are on the order of the Alfvenic transit times. The third term is the divergence of the electron pressure tensor, and the last term is the electron inertial term.

In fluid simulations, the electron pressure tensor is usually approximated by a scalar pressure, however it is known from fully kinetic simulations that off-diagonal terms can support reconnection electric fields. It has generally become clear that kinetic physics, being the first-principles description of a plasma, are important for a complete description of magnetic reconnection, in particular in the collisionless regime. Advances in computing power have enabled comparatively large fully kinetic simulations using the particle-in-cell (PIC) method, even in 3-d, that can achieve a good separation between the global size of the simulation and ion and electron scales, even though they usually require a reduced ion/electron mass ratio to keep the computational cost manageable. These kinds of simulations are limited to local simulations of reconnection, although their extent may be 10s or 100s of ion inertial scales ($d_i$). In contrast, many physical systems of interest, e.g., Earth's magnetosphere or the Sun's corona are many orders of magnitude larger and will remain inaccessible to fully kinetic simulations for the foreseeable future.

Fluid simulations are computationally significantly cheaper than kinetic simulations due to their reduced dimensionality, making them computationally more attractive but at the cost of a loss of kinetic information in the closure formulation.

An alternative fluid formulation employed in this context is the multi-fluid moment model. This model is derived without any approximations by taking moments of the Vlasov equation and keeps the full set of Maxwell's Equations.

\subsection{Multi-Fluid and Ten Moment Model}
\begin{equation} 
    d_t f_s = \partial_t f _s+  \Vec{v}\cdot \nabla_r f_s + \frac{q}{m}\left( \Vec{E} + \Vec{v}\times \Vec{B}\right) \cdot \nabla_v f_s = 0
    \label{eqn:vlasov}
\end{equation}
The multi-fluid model is constructed by taking sequentially increasing velocity space moments of the Vlasov equation~\eqref{eqn:vlasov} for each species $s$.  Following \citep{Wang2015}, the following moments are defined:
\begin{subequations}
    \label{eqn:allmomentdefs}
    \begin{gather}
    n\equiv \int f d\Vec{v} 
        \label{eqn:zeromoment} \\
    u_j \equiv \frac{1}{n} \int v_j f d\Vec{v}
        \label{eqn:firstmoment} \\
    \mathcal{P}_{ij}\equiv m\int v_iv_j f d\Vec{v}
        \label{eqn:secondmoment} \\
    \mathcal{Q}_{ijk}\equiv m\int v_iv_jv_k f d\Vec{v}
        \label{eqn:thirdmoment}
    \end{gather}
\end{subequations}
The evolution equations for these moments are derived by multiplying the Vlasov equation by consecutive powers of $v$ and integrating out velocity space. Truncating the system of equations after the 2nd order gives the so-called 10-moment equations~\eqref{eqn:allmomentequations}, which describe the evolution of density, the three components of momentum, and each of the 6 unique components of the symmetric pressure tensor. Each species is maintained as a separate set of fluid moments, but a species index has been left out above for brevity.

\begin{subequations}
    \label{eqn:allmomentequations}
    \begin{gather}
    \partial_t n + \partial_{j} (nu_j) = 0
        \label{eqn:zerothmomentequation} \\
    m\partial_t(nu_i) + \partial_{j}\mathcal{P}_{ij} = nq\left( E_i + \epsilon_{ijk}u_jB_k\right)
        \label{eqn:firstmomentequation} \\
    \partial_t \mathcal{P}_{ij} + \partial_{k} \mathcal{Q}_{ijk} = nqu_{[i}E_{j]} + \frac{q}{m}\epsilon_{[ikl}\mathcal{P}_{kj]}B_l
        \label{eqn:secondmomentequation}
    \end{gather}
\end{subequations}
The square bracket notation above indicates the minimal sum over free indices that yields a completely symmetric tensor.  The electromagnetic fields evolve according to the Maxwell equations
\begin{subequations}
    \begin{gather}
    \nabla \times \Vec{E} = -\frac{\partial\Vec{B}}{\partial t}\\
    \nabla \times \Vec{B} = \mu_0 \Vec{J} + \frac{1}{c^2}\frac{\partial\Vec{E}}{\partial t}
    \end{gather}
\end{subequations}

The above moments~\eqref{eqn:allmomentdefs} are used for the derivation of the ten moment model given by~\eqref{eqn:allmomentequations}. Frequently however, working with the centered moments~\eqref{eqn:allcenteredmomentdefs} is preferred because of their convenient physical interpretation as the pressure stress tensor and heat flux tensor.  
\begin{subequations}
    \label{eqn:allcenteredmomentdefs}
    \begin{gather}
        P_{ij} \equiv m\int (v_i - u_i)(v_j - u_j) f d\Vec{v} \label{eqn:secondcenteredmoment} \\
        Q_{ijk} \equiv m\int (v_i - u_i)(v_j - u_j)(v_k - u_k) f d\Vec{v} \label{eqn:thirdcenteredmoment}
    \end{gather}
\end{subequations}

The 10-moment model may be re-written in terms of the centered moments, which are related according to
\begin{subequations}
    \label{eqn:momentrelation}
    \begin{gather}
        \mathcal{P}_{ij} = P_{ij} + 2nmu_iu_j
            \label{eqn:prelation} \\
        \mathcal{Q}_{ijk} = Q_{ijk} + u_{[i}\mathcal{P}_{jk]} - 2nmu_iu_ju_k
            \label{eqn:qrelation}
    \end{gather}
\end{subequations}

Noticeably,  each moment equation contains the next higher order moment.  This trend continues ad-infinitum resulting in an open system of equations.  
While this system of equations is exact, to be of practical use it needs to be truncated. We truncate the 10-moment model by seeking a closure that replaces $Q$ with an expression containing known quantities, such as the lower order moments.  In the adiabatic case, $P_{ij}$ is taken to be isotropic and $Q$ is zeroed out, giving the 5-moment model. In this work, we are specifically interested in the pressure tensor's impact on reconnection and hence choose to retain its full structure.

Previous closures have been proposed such as the CGL family of closures~\citep{Chust2006} and the Hammett-Perkins closure~\citep{Hammet1990}.

\textbf{In this work we aim to use data-driven methodologies to derive a symbolic closure relation truncating the cascading system of equations at the 10-moment model.}  

We build off the work of \citep{Wang2015}, investigating the collisionless regime using data from a fully kinetic Harris Sheet PIC simulation.  We then apply the data-driven technique, Sparse Identification of Nonlinear Dynamics (SINDy)~\citep{Brunton2016}, to distill the raw particle output into a symbolic closure which is compared to the approximate local closure~\eqref{eqn:wang_closure}
\begin{equation}
    \partial_m Q_{ijm} \approx v_t |k_0|\left(P_{ij} - p\delta_{ij} \right)
    \label{eqn:wang_closure}
\end{equation}
In the above, $v_t$ is the thermal velocity, $k_0$ is a typical scale defining wave number, $p$ is the scalar pressure attained by averaging the diagonal of the pressure stress tensor, and the moments are centered. Originally defined in~\citep{Wang2015}, \eqref{eqn:wang_closure} is referred to as the local approximate Hammett-Perkins closure because it replaces the continuous $k$ in the general non-local Hammett-Perkins closure, with a single $k_0$ value in physical space.

In the literature various options for closures to the 10-moment model have been investigated~\citep{Ng2017,Ng2020} by implementing a given proposed closure into the multi-fluid moment model, choosing parameters like a typical wavelength of kinetic instability $k_0$ and running prototypical reconnection problems like the Harris sheet or island coalescence. The time evolution and snapshots of the fields are then compared to results of a particle-in-cell simulation.

In this work on the other hand, we exploit the additional information available from PIC simulations to test and derive closures. Since we do have the fully kinetic particle information available, we can for example calculate the third order heat flux moment and compare the actual heat flux observed in the simulation to the assumptions of a given closure. This process has the advantage that we can directly compare terms at any given time, whereas when comparing multi-fluid and PIC simulations, the state at any given time is an accumulation of all differences that have occurred up to this time.

\section{Background}\label{sec:background}
\subsection{Particle-In-Cell Method}
The Particle Simulation Code (PSC)~\citep{Germaschewski2016} is a modern, load-balanced, GPU accelerated, fully-kinetic particle-in-cell code.  The PIC method is a numerical method which discretizes the electromagnetic fields on a 3D grid and advances them using Maxwell's equations while approximating the particle distribution function through macroparticles which are advanced in continuous phase space. 

As a fully kinetic algorithm, the PIC method solves the full Vlasov-Maxwell system of equations.  Thus moments of the distribution function evolved by the PIC method exactly satisfy the 10-moment model up to noise and other numerical errors.  Systemic noise inherent to the PIC method is introduced primarily through the sampling of macroparticles to construct distribution functions.  

This systemic noise was demonstrated to be a significant problem when applying the SINDy method to PIC data by~\citep{Alves2020}.  By spatially integrating the PIC data, they were able to reduce noise and apply SINDy to effectively recover the Vlasov equation, recover fluid equations, and discover an adiabatic closure.  In~\S\ref{sec:experiments} we borrow this technique and apply it to each of the moments and fields output by PSC before any further operation is performed.

%The moments calculated by PSC~\eqref{eqn:allmomentdefs} retain the relevant kinetic physics and satisfy the 10-moment model which we will demonstrate in~\S\ref{sec:10moment verification} . 

\subsection{Sparse Identification of Non-Linear Dynamics}
The Sparse Identification of Non-Linear Dynamics is a framework for the discovery of symbolic equations.  Developed by~\citep{Brunton2016}, it looks to construct parsimonious solutions from a library containing candidate terms which are calculated from raw data.  It may be used to take data from either experiment or simulation and generate symbolic governing equations which describe a dynamical system. 

This is accomplished by solving the sparsity-promoting regression problem
\begin{equation}
\mathcal{L} = \min_\Vec{w} ||\Theta\Vec{w} - \Vec{y} ||_2 + \lambda R(\Vec{w})
\label{eqn:lassoregression}
\end{equation}
Where $\Theta$ is the library containing candidate terms, $\Vec{w}$ is the vector of coefficients associated with each term in the library, $\Vec{y}$ is the regression target, and $R(\Vec{w})$ is some regularization function on $\Vec{w}$ scaled by constant $\lambda$.  Commonly $R$ is the $L_1$-norm in which case the above becomes Lasso regression. For a detailed description of the framework, see~\citep{Zheng2019,Champion2020}.

Our experiments approximate~\eqref{eqn:lassoregression} by solving the least squares problem with thresholding.  This involves iteratively solving the least squares problem $y=\Theta\Vec{w}$ and applying an upper and lower bounded threshold to the discovered $\Vec{w}$, thus restricting the coefficients to a pre-specified range. 

The success of SINDy discovering a parsimonious equation is dependent on selecting an upper and lower bounds which maximally encourages sparse solutions while minimally allowing error.  Solutions to problems of this sort are known as pareto-optimal if the family of solutions can not improve on one measure without hindering the other.  In other-words, the optimal bounds for the thresholded least squares problem are associated with the solution that lives on the pareto-front where the number of included terms may not be further minimized without increasing the solution error.

Thus, for each instance in which we apply SINDy, we first solve the thresholded least squares problem with various bounds to construct the pareto curve.  The optimal bounds are then selected from the pareto-front and the problem is re-solved with those bounds.  

This method was selected to symbolically model our closure because it is fully interpretable, emulates nature by leveraging parsimony, and has previously been used successfully to model both kinetic and fluid plasma systems~\citep{Alves2020,Kaptanoglu2021}.

\section{Experiments}\label{sec:experiments}

\subsection{Harris Sheet Simulation}
This study uses the PSC to simulate a collisionless Harris sheet reconnection problem. Initially in kinetic equilibrium, the magnetic field is initialized as $\Vec{B} = B_0\tanh(y/\lambda_B)\hat{\Vec{x}}$ and the densities as $n_e=n_i=n_0\sech^2(y/\lambda_B) + n_b$.  The magnetic field is then given a small sinusoidal perturbation which initiates the reconnection process.

The run uses a $1280 \times 640$ grid with a mass ratio of 25 and $16\times10^9$ particles which resolves to 0.676 $\Delta x$/Debye length.

This simulation uses an abnormally high number of particles per cell for each species, approximately 10,000.  This was done to lower the level of noise inherent to PIC methods as demonstrated by~\citep{Juno2020}.

\begin{table}
  \begin{center}
  \begin{tabular}{cccccc}
      $L_x/d_{i0}$  & $L_y/d_{i0}$   &   $L_z/d_{i0}$ & $n_b/n_0$ & $T_{i0}/T_{e0}$ & $\lambda_B/d_{i0}$  \\[3pt] \hline
        25 & 1 & 12.5 & 0.3 & 5 & 0.5 \\
  \end{tabular}
  \caption{Harris Sheet Simulation Numerical Parameters}
  \label{tab:}
  \end{center}
\end{table}

\subsection{Method}
Our method, which relies on SINDy at its core, was used for verifying the ten moment model, analysing the existing local approximate Hammett-Perkins closure, and searching for an improved closure. 

We begin by loading the moment and field data generated by PSC and calculating relevant spatial and temporal derivatives.  We then spatially integrate each term and construct a library of terms consisting of every r.h.s term from the 10-moment model.

In verifying the ten moment model, each order's term library $\Theta$ is constructed by selecting out only terms with consistent units.  Specifically, the 0th order's term library $\Theta$, contains only two terms from from the single 0th order equation (reducing the method to standard regression), the 1st order's term library contains 9 possible terms from the three 1st order equations, and the 2nd order's term library contains 24 possible terms from the six 2nd order equations.  Pairing terms with consistent physical units acts as an a-priori constraint on which terms may be accepted into the library. 

With the constructed library we then calculate the l.h.s of each equation in the 10-moment model.  For consistency, we consider the l.h.s to be the derivative containing terms of the equation, with the exception of the 0th order where the partial time derivative is the l.h.s and the particle divergence is the r.h.s. 

The l.h.s is then used as a regression target wherein we apply the SINDy framework on the associated term library to discover the associated symbolic equations constituting the r.h.s.  This method is summarized in~\cref{alg:1}.

\begin{algorithm}
\caption{Equation Synthesis}
\begin{algorithmic}[1]
\State $E, B, f^i \gets \text{Load field and moment data}$
\State Calculate $\partial_tf^i$, $\partial_xf^i$, $\partial_yf^i$, $\partial_zf^i$ for relevant moments
\State Integrate and construct term library $\Theta$
\State define l.h.s as regression target
\State Apply SINDy with pareto-optimal bounds\\
\Return symbolic equation
\end{algorithmic}
\label{alg:1}
\end{algorithm}

\subsection{Ten Moment Model and Method Verification}\label{sec:10moment verification}
As the multi-fluid moment equations~\eqref{eqn:allmomentequations} are exact, we can verify them directly from the Harris sheet PIC data.  To show this we individually calculate the left hand side (l.h.s) and right hand side (r.h.s) of each equation and demonstrate their equivalence.

In the following analysis we use the antisymmetric normalized $L_2$ error~\eqref{eqn:norm_l2_error} as a method for four separate comparisons.  The first is the numerical error between the l.h.s and r.h.s calculated directly from the PIC simulation data.  This metric is used to verify that the kinetic data satisfies the 10-moment model and gives an indicator of the residual noise in the data after integration.  The second comparison is the numerical error between the r.h.s calculated directly from the data and the discovered r.h.s equation applied to the data.  This is an indicator for how well our method is able to reproduce theory.  The third comparison is the numerical error between the l.h.s calculated from the data and the discovered r.h.s equation applied to the data, which gives an indicator for how well our discovered equation can explain the l.h.s of each moment relation.  

The final comparison we make is the normalized $L_2$ metric between the coefficients given by the known multi-fluid equations~\eqref{eqn:allmomentequations} and those discovered using the regression method.  This is referred to as the coefficient error.   
\begin{equation}
   L_2(\Vec{x}_1,\Vec{x}_2)= \frac{||\Vec{x}_1 - \Vec{x}_2||_2}{||\Vec{x}_1||_2}
   \label{eqn:norm_l2_error}
\end{equation}

Using this analysis we find that the regression method is able to reproduce the multi-fluid equations if the $L_2$(l.h.s, discovered) error is equal to or approximately equal to the $L_2$(l.h.s, r.h.s) error.  Ultimately we are looking for symbolic, not numeric closure approximations.  With this in mind the coefficient error is the truest indicator for the success of the method in reproducing~\eqref{eqn:allmomentequations}. 

\begin{figure*}
    \centering
        \includegraphics[width=.75\linewidth]{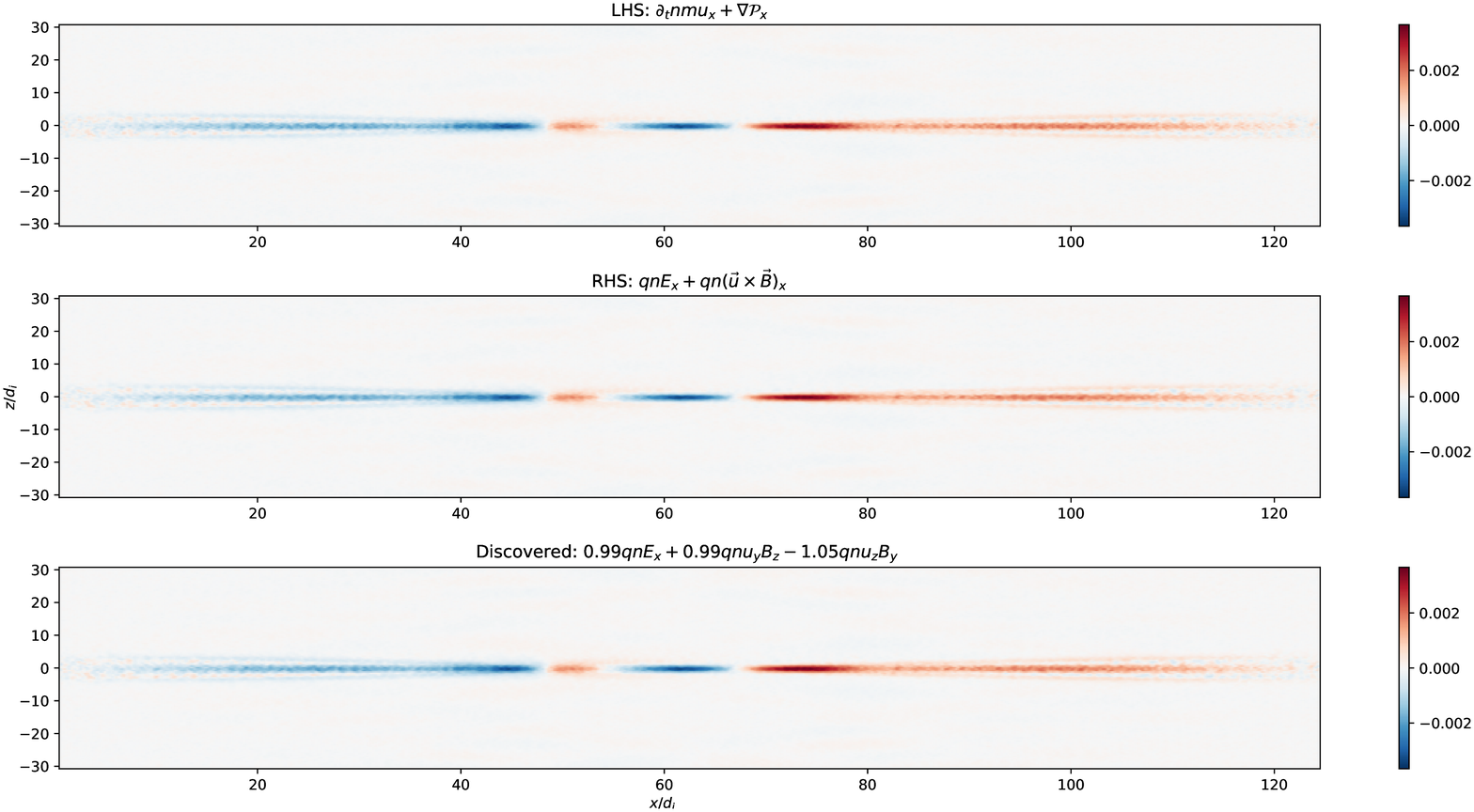}\par 
        \includegraphics[width=.75\linewidth]{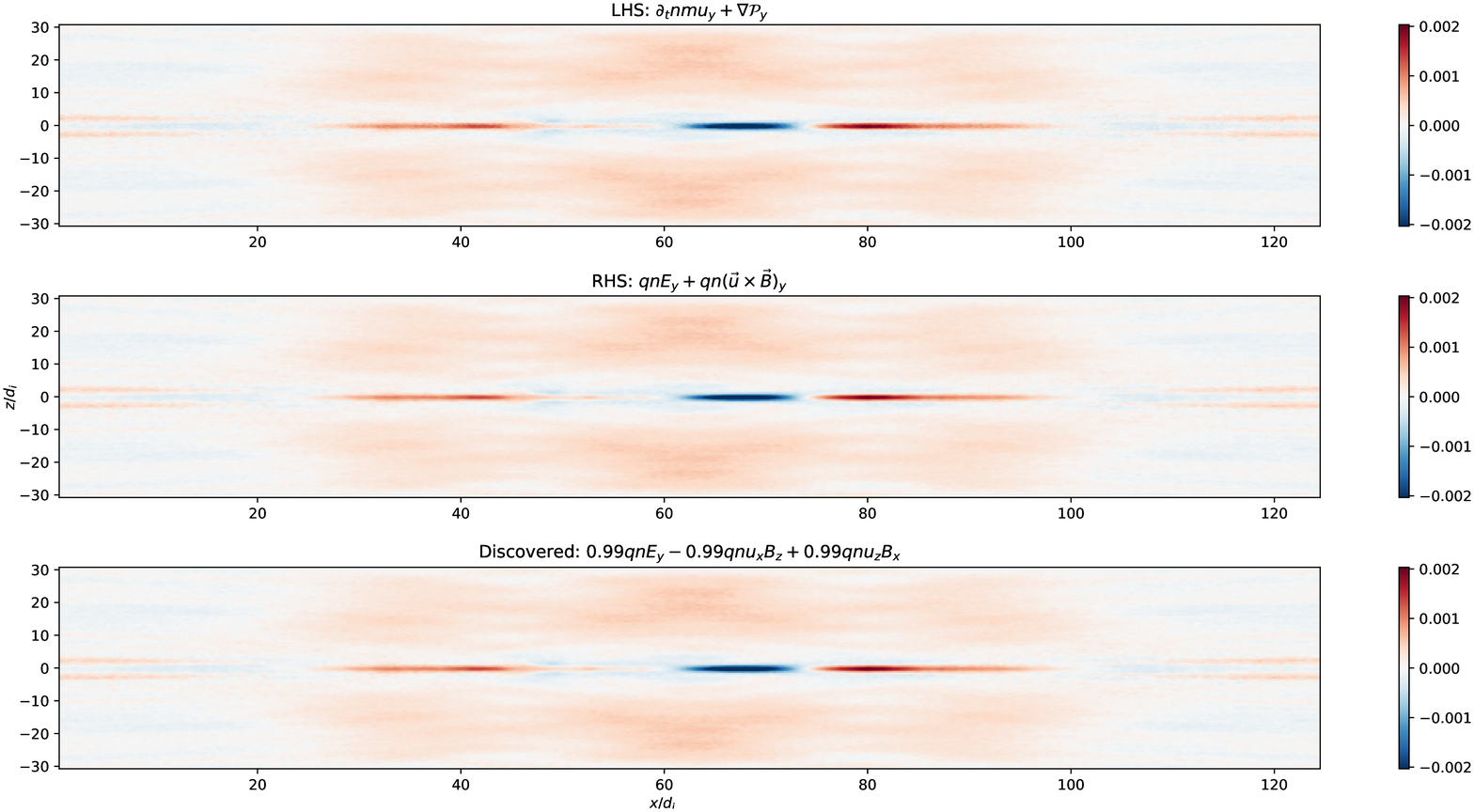}\par 
        \includegraphics[width=.75\linewidth]{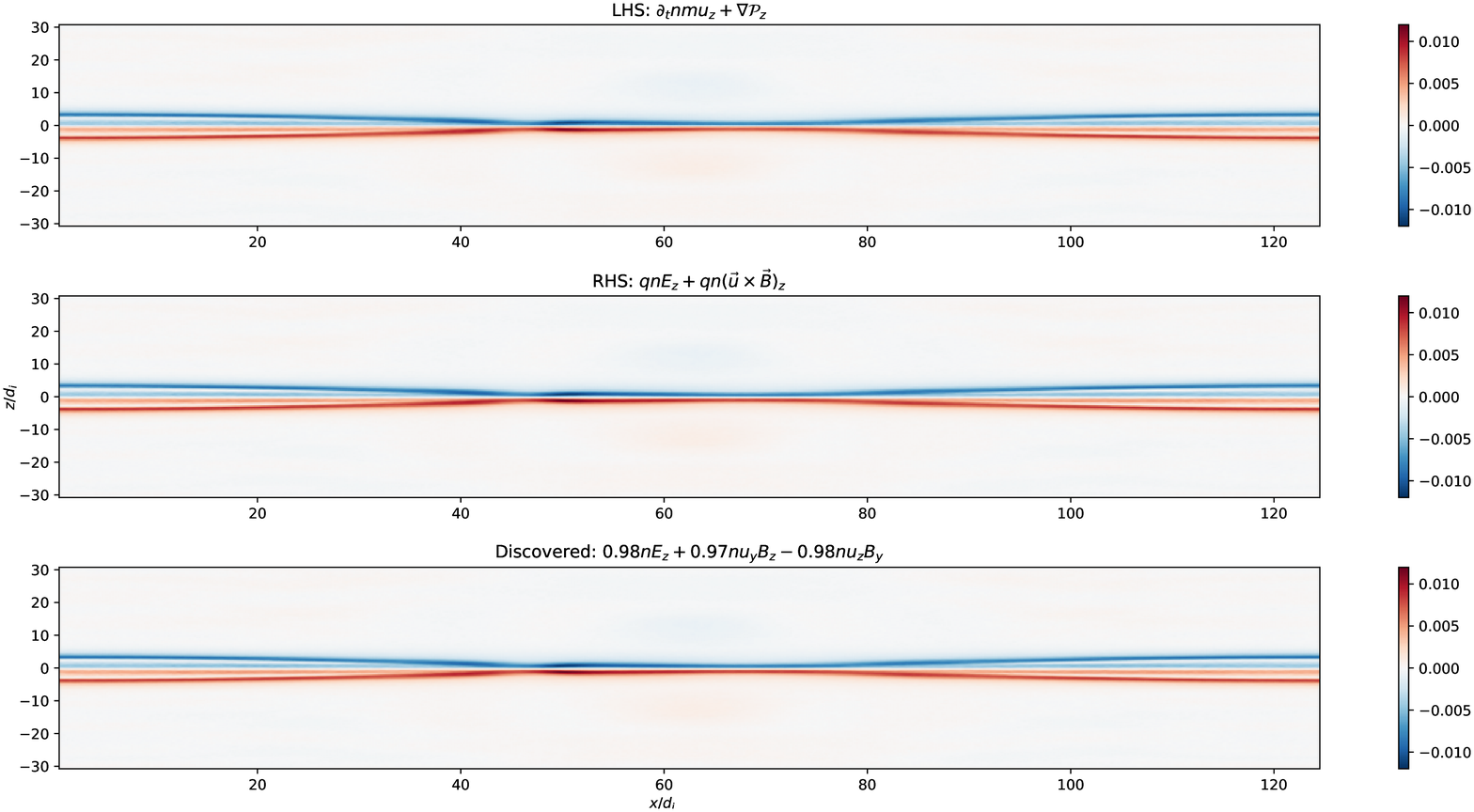}\par 
    \caption{Top to bottom are the x-component, y-component, and z-component of the 1st order (momentum) equation with the l.h.s in the top pane, the true r.h.s in the middle pane, and the discovered r.h.s in the bottom pane}
    \label{fig:1st_verification}
\end{figure*}

\begin{figure*}
    \centering
    \begin{multicols}{2}
        \includegraphics[width=\linewidth]{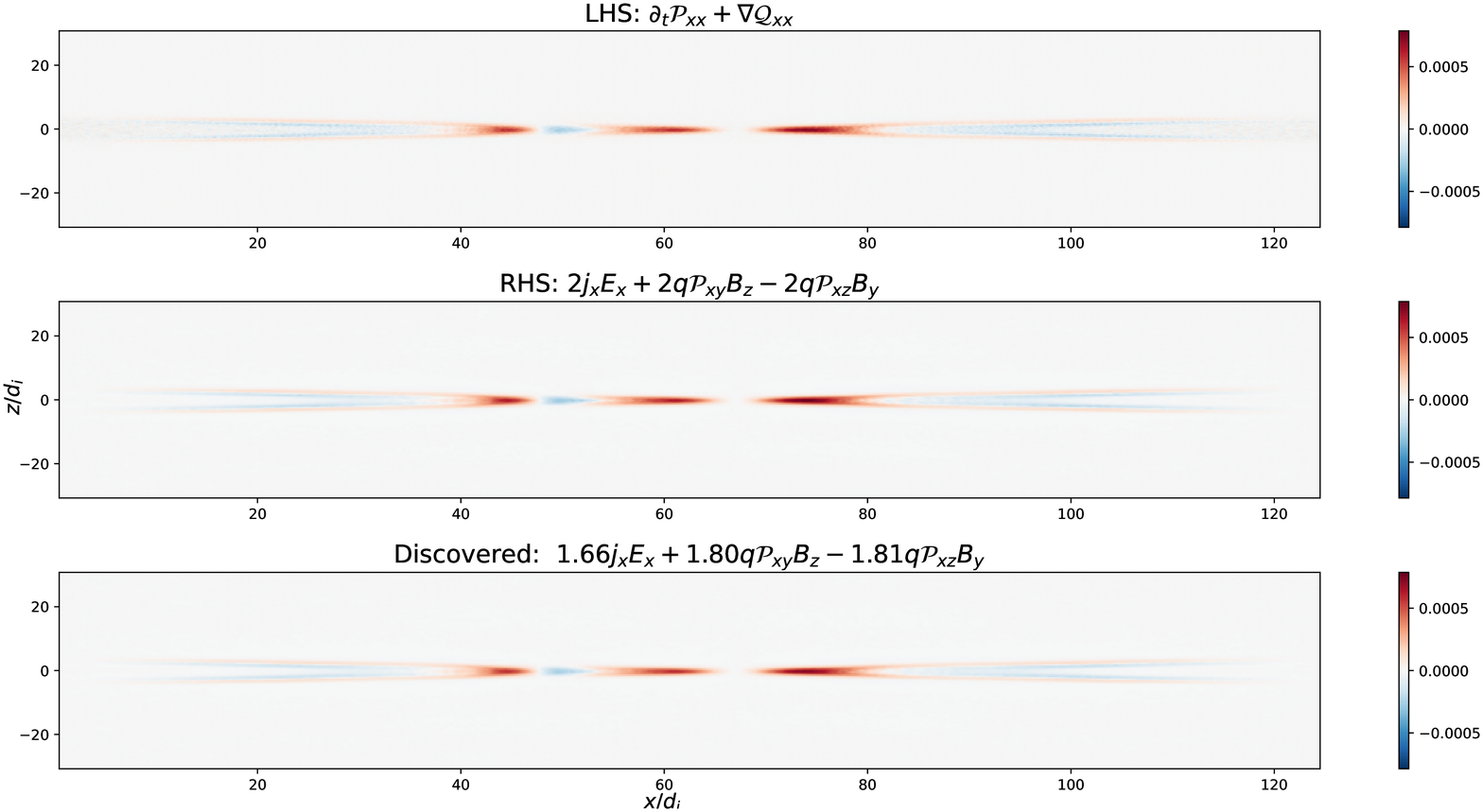}\par 
        \includegraphics[width=\linewidth]{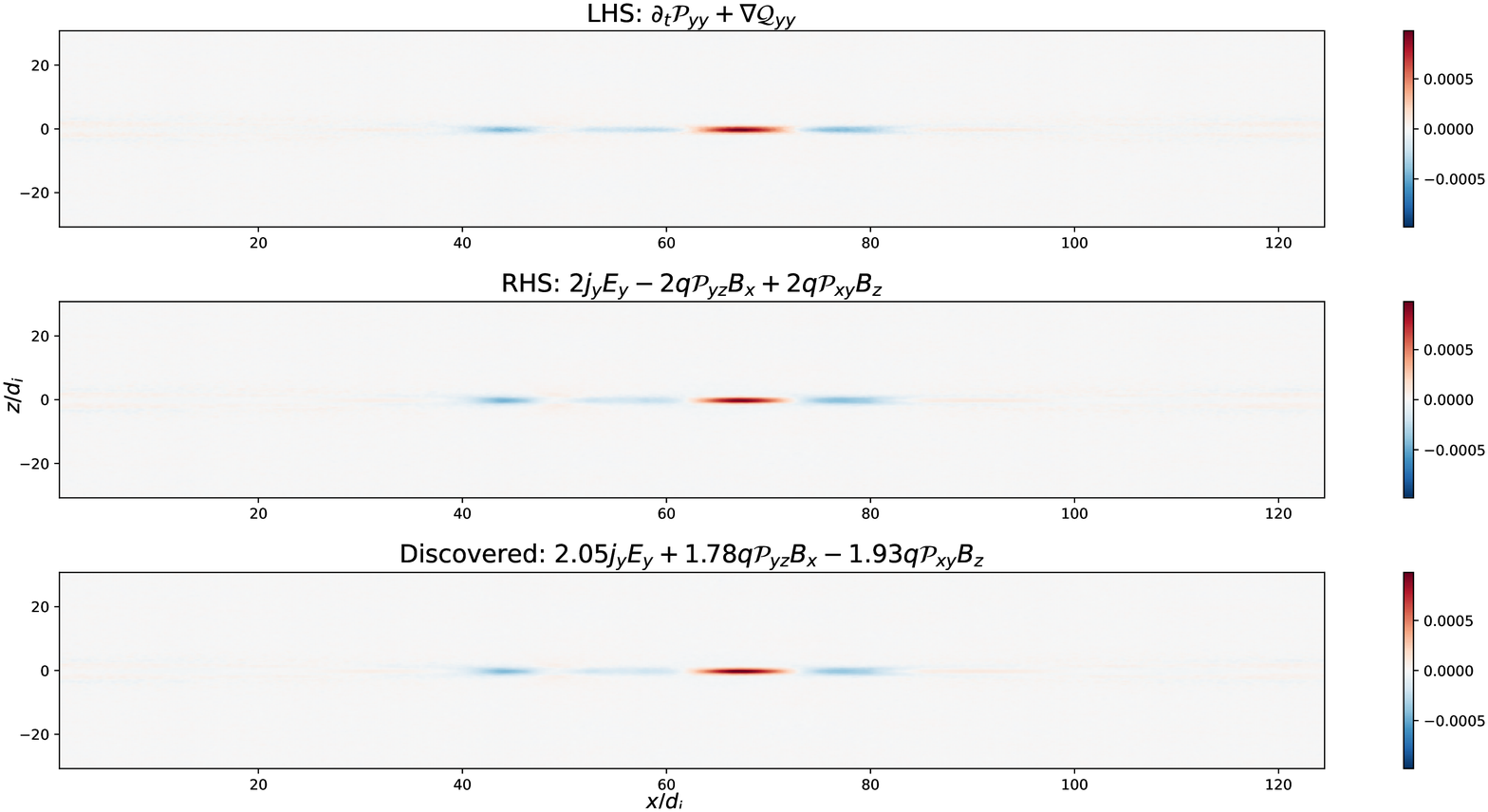}\par 
    \end{multicols}
    \begin{multicols}{2}
        \includegraphics[width=\linewidth]{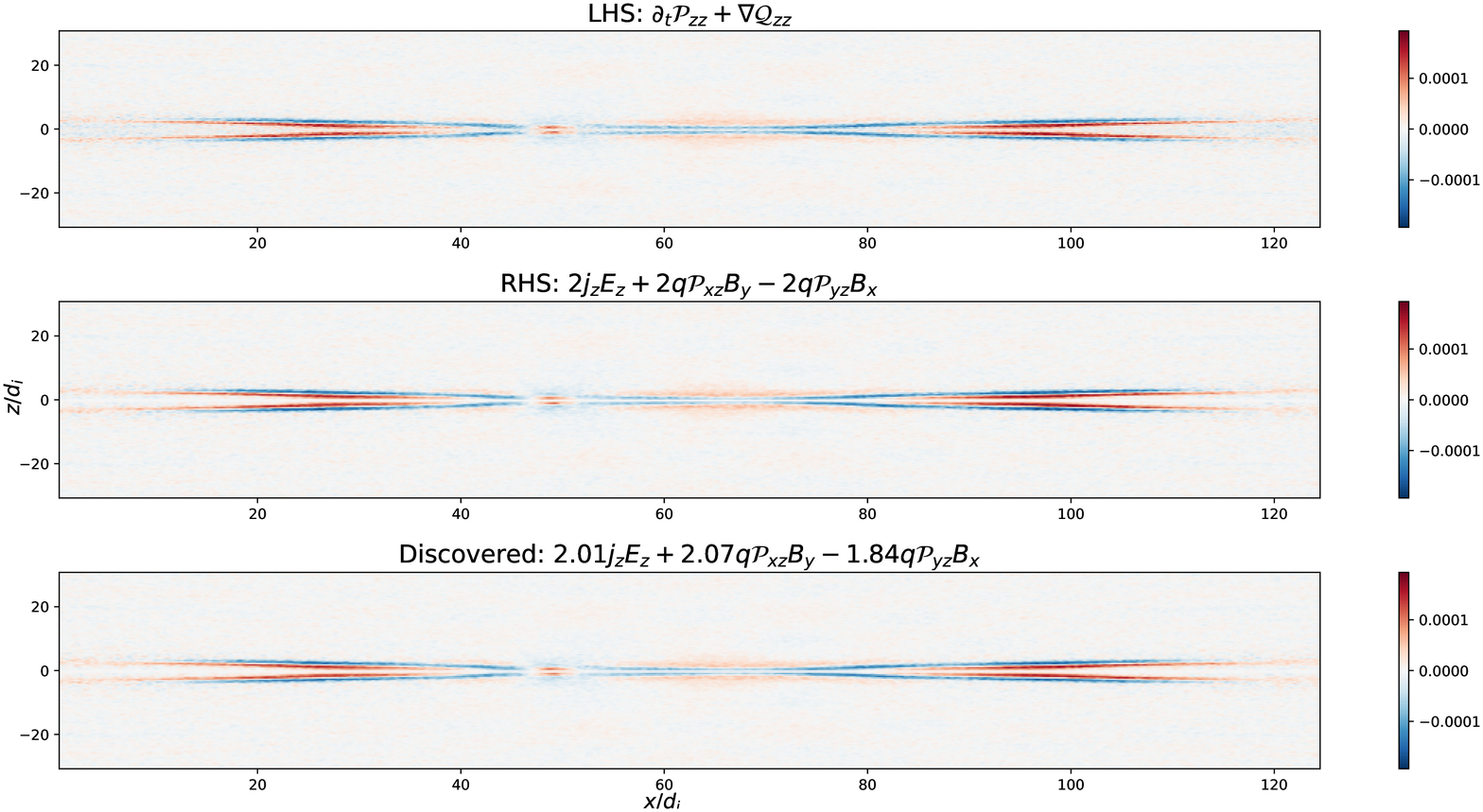}\par 
        \includegraphics[width=\linewidth]{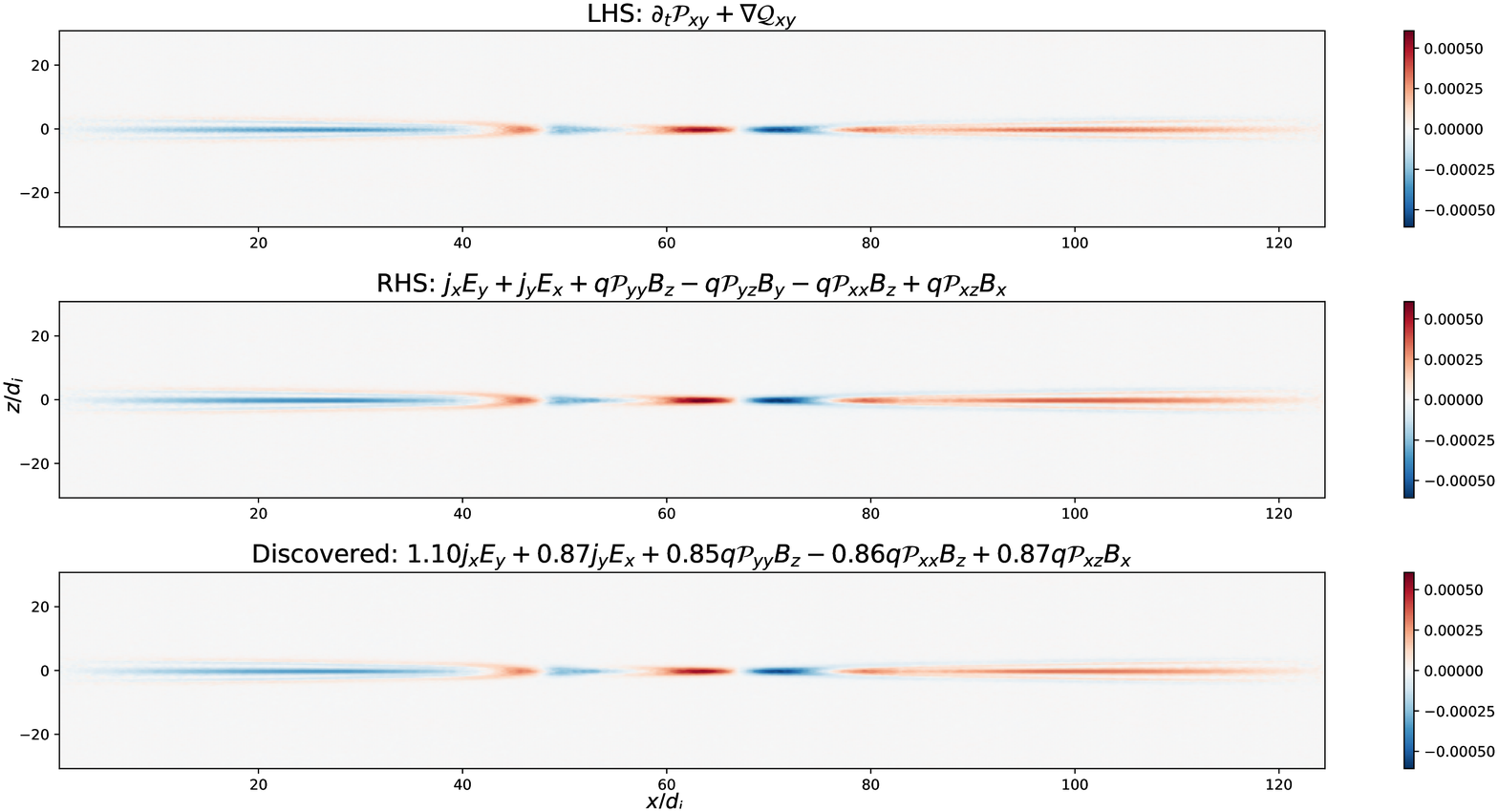}\par 
    \end{multicols}
    \begin{multicols}{2}
        \includegraphics[width=\linewidth]{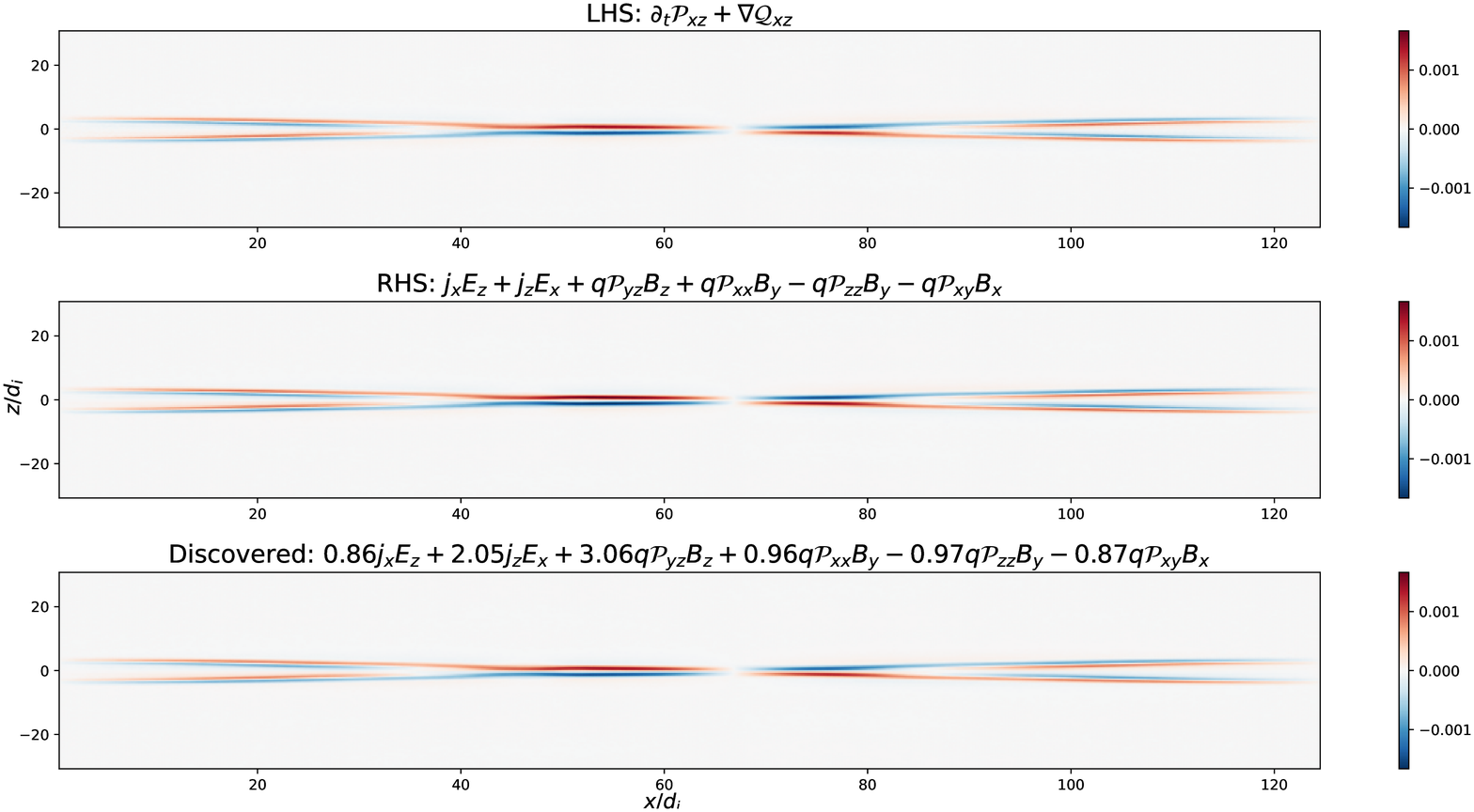}\par 
        \includegraphics[width=\linewidth]{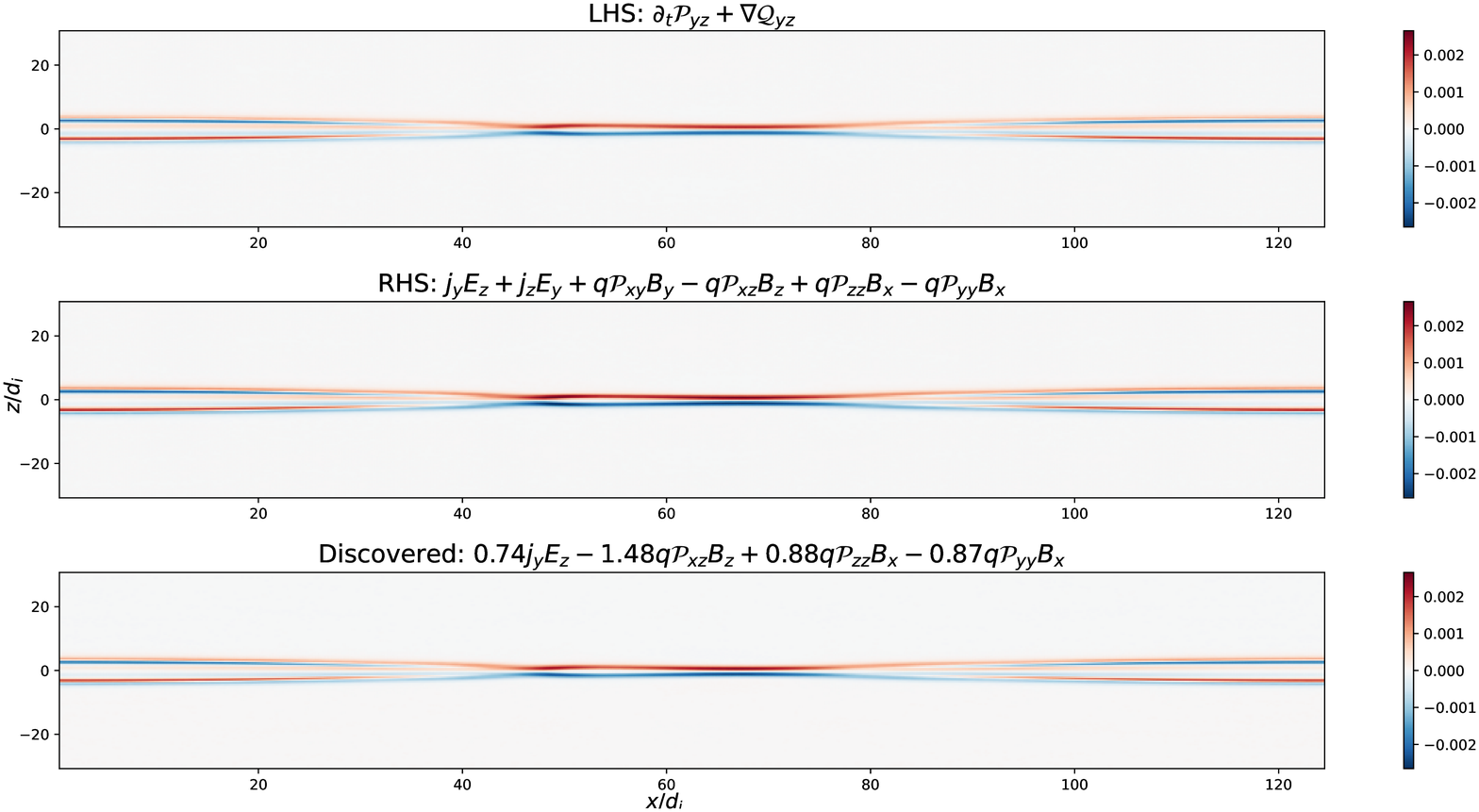}\par 
    \end{multicols}
\caption{Top to bottom, left to right are the xx, yy, zz, xy, xz, yz components of the 2nd order moment equation.  Each contains the regression target in the top pane, the true r.h.s in the center pane, and the discovered relation in the bottom.   Visually, the l.h.s, r.h.s, and discovered source term of each component match well and are consistent with theory. This is true of both the on-diagonal and off-diagonal terms which demonstrates the reliability of our data.  While some off-diagonal discovered terms may be missing, this is most likely due to their magnitude being small in this particular scenario.}
\label{fig:2nd_verification}
\end{figure*}

We present our findings in~\cref{tab:0thresults,tab:1stresults,tab:2ndresults} and~\cref{fig:0th_verification,fig:1st_verification,fig:2nd_verification}, where the data presented is from a representative step T-17 $\omega_{ci}^{-1}$ (ion cyclotron frequency), during the nonlinear phase of reconnection. Experiments were run at various time steps throughout the nonlinear phase and the results were shown to not differ greatly.

Systematic error is introduced into the analysis when approximating the time and space derivatives with the centered finite-difference method.  This error appears in the $L_2$(l.h.s, r.h.s) and the $L_2$(l.h.s, discovered) errors presented in the analysis and does not indicate a violation of the conservation laws.  

There is potential that this source of error could pose a problem in discovery of derivative containing terms.  So while this wouldn't be a problem for the method verification of \S\ref{sec:10moment verification} where the term library $\Theta$ does not contain derivative terms, it could raise issues for closure discovery in \S\ref{sec:closure_discovery}. To investigate whether this error was a problem, the method verification of \S\ref{sec:10moment verification} was repeated for each equation using only the partial time derivative term as the regression target.  This modification required the spatial derivative terms be included in $\Theta$.  Results showed that noise induced by the spatial derivatives did not pose a problem in r.h.s discovery. The derivative terms were found with error on the order of results as presented. 

With semantic consistency in mind, we thus stick to the convention of the 10-moment model and use the derivative containing terms as the l.h.s throughout the following sections.

%similarly low error, justifying our use of the derivative containing terms as the l.h.s for semantic consistency. 

\subsubsection{0th Order Verification}
%%%%%%%%%%%%% 0th order
\begin{table}
\centering
\resizebox{\textwidth}{!}{\begin{tabular}{|c|c|c|c|c|c|}\hline
    r.h.s & discovered & $L_2$(l.h.s, r.h.s) & $L_2$(r.h.s, discovered) & $L_2$(l.h.s, discovered) & coefficient error\\ \hline
    $-\nabla\cdot(n\vec{u})$ & $-1.06\partial_xnu_x - 1.07\partial_znu_z$ & 0.234 & 0.071 & 0.225 & 0.065 \\ \hline

\end{tabular}}
\caption{The true r.h.s, discovered r.h.s, and systematic errors of the 0th order moment equation.}
\label{tab:0thresults}
\end{table}

\begin{figure}
    \centering
    \includegraphics[width=\linewidth]{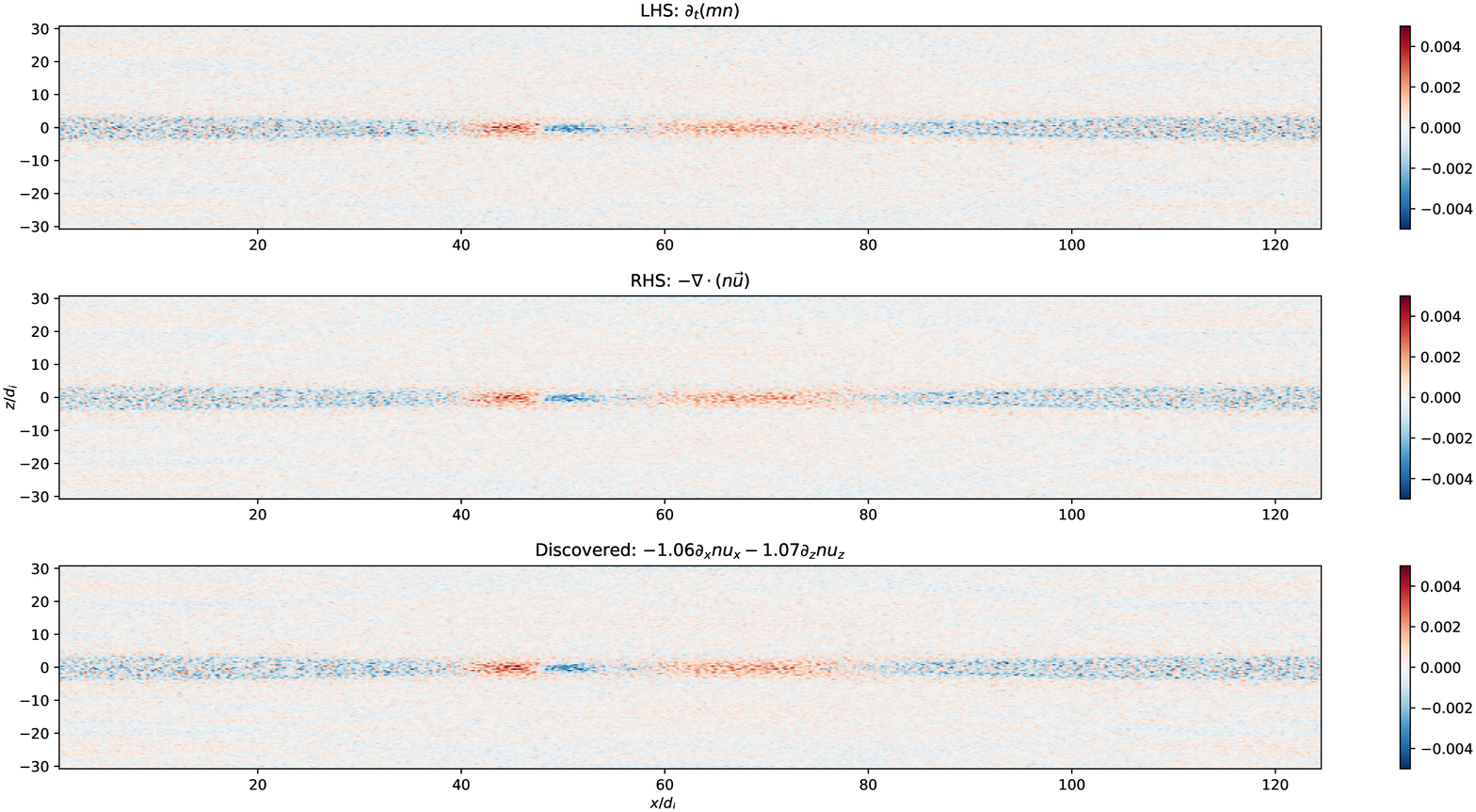}
    \caption{The 0th order moment equation or the continuity equation verified from the Harris sheet PIC data. Here we contrast the l.h.s in the top pane with the true r.h.s in the center pane and the discovered r.h.s in the bottom pane}
    \label{fig:0th_verification}
\end{figure}

The 0th order equation, calculated from kinetic data, is verified visually as seen by the close matching in~\cref{fig:0th_verification} and the low $L_2$(l.h.s, r.h.s) error.  One might believe that the residual noise would distort the ability of SINDy to isolate the correct dynamics.  This however is not the case as demonstrated by the low coefficient error, 0.065. This demonstrates the robustness of this method to residual noise and the derivative approximation systematic error.   This is important for validating the approach for use in closure discovery where many library terms will contain derivatives.

% Noise generated in the centered finite-difference derivatives introduce systematic error that leads to error in the regression. 

\subsubsection{1st Order Verification}
%%%%%%%%%% 1st order
\begin{table}
\centering
\resizebox{\textwidth}{!}{\begin{tabular}{|c|c|c|c|c|c|c|}

    component & r.h.s & discovered & $L_2$(l.h.s, r.h.s) & $L_2$(r.h.s, discovered) & $L_2$(l.h.s, discovered) & coefficient error \\ \hline
    x & $qnE_x + qn(\vec{u} \times \vec{B})_x$ & $0.99qnE_x + 0.99qnu_yB_z -1.05qnu_zB_y$ & 0.113 & 0.008 & 0.113 & 0.030\\ \hline
    
    y & $qnE_y + qn(\vec{u} \times \vec{B})_y$ & $0.99qnE_y - 0.99qnu_xB_z + 0.99qnu_zB_x$ & 0.074 & 0.007 & 0.074 & 0.010 \\ \hline
    
    z & $qnE_z + qn(\vec{u} \times \vec{B})_z$ & $0.98nE_z + 0.97nu_yB_z - 0.98nu_zB_y$ & 0.044 & 0.014 & 0.041 & 0.023\\ \hline
    
\end{tabular}}
\caption{The true r.h.s, discovered r.h.s, and systematic errors for each component of the 1st order moment equations.}
\label{tab:1stresults}
\end{table}

The inclusion of field terms in the 1st order equations stabilizes the numerics significantly.  This is demonstrated by the superb l.h.s-r.h.s matching, the low $L_2$(l.h.s, r.h.s) error, and the low $L_2$(l.h.s, discovered) error for each of the three components.  Similarly, from the coefficient error we see that the method was able to reproduce the 1st order equations with 0.01-0.03 normalized $L_2$ coefficient error.  These are strong results which validate both the use of kinetic data for use in the 10-moment model and the use of SINDy to discover dynamics using the data.

\subsubsection{2nd Order Verification}
%%%%%%%%%%% 2nd order
\begin{table}
\centering
\resizebox{\textwidth}{!}{\begin{tabular}{|c|c|c|c|c|c|c|}

    component & r.h.s & discovered & $L_2$(l.h.s, r.h.s) & $L_2$(r.h.s, discovered) & $L_2$(l.h.s, discovered) & coefficient error \\ \hline
    xx &  $2j_xE_x + 2q\mathcal{P}_{xy}B_z - 2q\mathcal{P}_{xz}B_y$ & $1.66j_xE_x + 1.80q\mathcal{P}_{xy}B_z - 1.81q\mathcal{P}_{xz}B_y$ & 0.215 & 0.084 & 0.195 & 0.126 \\ \hline
    
    yy & $2j_y E_y - 2q\mathcal{P}_{yz}B_x + 2q\mathcal{P}_{xy}B_z$ & $2.05j_y E_y - 1.78q\mathcal{P}_{yz}B_x + 1.93q\mathcal{P}_{xy}B_z$ & 0.213 & 0.110 & 0.183 & 0.068\\ \hline
    
    zz & $2j_z E_z + 2q\mathcal{P}_{xz}B_y - 2q\mathcal{P}_{yz}B_x$ &  $2.01j_z E_z + 2.07q\mathcal{P}_{xz}B_y - 1.84q\mathcal{P}_{yz}B_x$ & 0.288 & 0.093 & 0.272 & 0.050\\ \hline
    
    xy &$j_xE_y + j_yE_x +q\mathcal{P}_{yy}B_z - q\mathcal{P}_{yz}B_y - q\mathcal{P}_{xx}B_z + q\mathcal{P}_{xz}B_x$ &  $1.10j_xE_y + 0.87j_yE_x + 0.85q\mathcal{P}_{yy}B_z - 0.86q\mathcal{P}_{xx}B_z + 0.87q\mathcal{P}_{xz}B_x$& 0.174 & 0.122 & 0.107 & 0.425\\ \hline
    
    xz & $j_xE_z + j_zE_x + q\mathcal{P}_{yz}B_z + q\mathcal{P}_{xx}B_y - q\mathcal{P}_{zz}B_y - q\mathcal{P}_{xy}B_x$ & $0.86j_xE_z + 2.05j_zE_x + 3.06q\mathcal{P}_{yz}B_z + 0.96q\mathcal{P}_{xx}B_y - 0.97q\mathcal{P}_{zz}B_y - 0.87q\mathcal{P}_{xy}B_x$ & 0.128 & 0.106 & 0.053 & 0.947 \\ \hline
    
    yz & $j_yE_z + j_zE_y + q\mathcal{P}_{xy}B_y - q\mathcal{P}_{xz}B_z + q\mathcal{P}_{zz}B_x - q\mathcal{P}_{yy}B_x$ & $0.74j_yE_z  - 1.48q\mathcal{P}_{xz}B_z + 0.88q\mathcal{P}_{zz}B_x - 0.87q\mathcal{P}_{yy}B_x$ & 0.212 & 0.151 & 0.123 & 0.623 \\ \hline
\end{tabular}}
\caption{The true r.h.s, discovered r.h.s, and systematic errors for each component of the 2nd order moment equations.}
\label{tab:2ndresults}
\end{table}
%- 0.00q\mathcal{P}_{yz}B_y  xx
%+ 0.00j_zE_y + 0.00q\mathcal{P}_{xy}B_y zz

The 2nd order equations demonstrate different behavior on-diagonal vs off-diagonal. The $L_2$(l.h.s, r.h.s) numerical errors are low in both the on-diagonal and off-diagonal terms, validating the 2nd order equations of the 10-moment model.  Similarly the low $L_2$(r.h.s, discovered) in conjunction with the low $L_2$(l.h.s, discovered) numerical error indicates that the terms included in the discovered solution account for the major contributions to the true r.h.s for each component.  This similarly implies that the terms excluded from the discovered r.h.s must be of small magnitude.  The small magnitude of these terms leads to discrepancies in the discovered r.h.s of the xy, xz, and yz directions, where the coefficient error is $> 0.4$.  This high coefficient error is due to the exclusion of terms from the discovered solutions.  Any term with a small magnitude will be eliminated by SINDy and is a constraint of the method.  Thus this inconsistency is explainable and expected in the case of minorly contributing terms.

% PIC methods contain intrinsic noise due to both particle resampling and mapping the particle's continuous position to discrete space. The higher order moments thus contain more particle noise which becomes amplified after subtracting out the lower order moments in order to obtain the centered moments.

\subsection{Local approximate Hammett-Perkins Closure}

% Another comment for (possibly) later. I think it might have been clearer to organize the paper by first attempting to verify the H-P closure, and then using the mismatch to motivate the ML attempt of discovering a better closure.

\begin{figure}
    \centering
        \includegraphics[width=\linewidth]{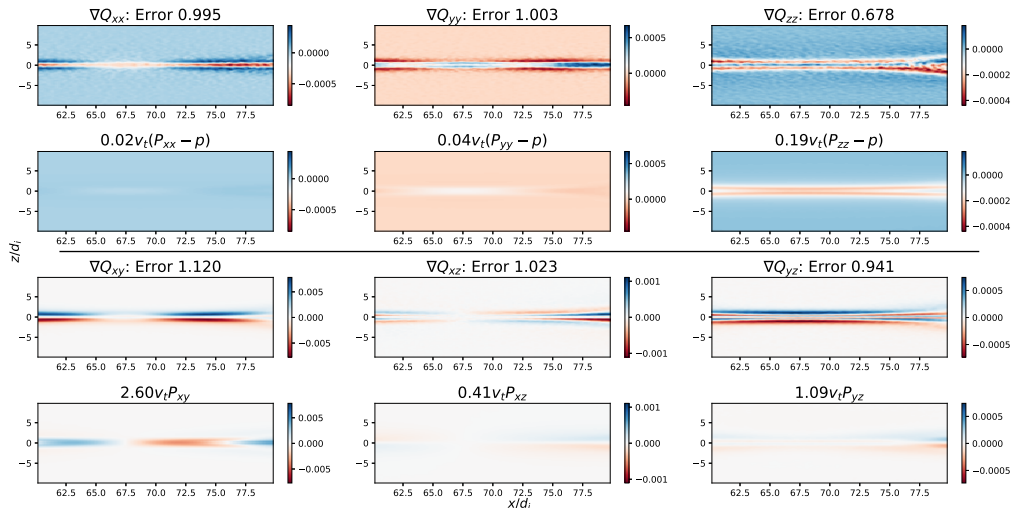}
    \caption{The local approximate Hammett-Perkins closure calculated during the nonlinear phase of reconnection.  In each coordinate, the top row shows the divergence of the heat flux, while the bottom row represents the prediction of the approximate closure.  The unknown factor $k_0$ for each component has been calculated by averaging $\frac{\nabla Q_{ij}}{v_t(P_{ij} - p\delta_{ij})}$ across the entire domain.}
    \label{fig:closure_verification}
\end{figure}

The local closure~\eqref{eqn:wang_closure} replaces the continuous wave-number $k$ in the Fourier space response with just one typical wave-number $k_0$.  Thus, omitting $k_0$ when calculating the closure we would expect the l.h.s and r.h.s to differ by a constant factor.  Further the closure is most important near the reconnection x-point where we expect important departures from ideal behavior.  Generally, these trends seem to be verifiable in the  kinetic data seen in~\cref{fig:closure_verification}, there is however room for improvement in several areas.  One obvious discrepancy is that each direction of the heat-flux divergence tensor scales by a different $k_0$ from the closure.  Second is the clear disparity between the heat-flux divergence and the closure at regions of the domain distant from the x-point.

Each component presented in~\cref{fig:closure_verification} is represented with $k_0$ calculated by taking the average $\frac{\nabla Q_{ij}}{v_t(P_{ij} - p\delta_{ij})}$ across the entire domain.  In all cases this results in the pressure terms being washed out with a lower magnitude than the heat flux divergence.  The on-diagonal terms appear to match structurally while the off-diagonal terms differ.  $\nabla Q_{xy}$ has some quadrupolar structure while the pressure terms appear to be bipolar.  The large scale structures of $\nabla Q_{xz}$ and $\nabla Q_{yz}$ are represented in the associated pressure terms, but the fine scale structures are omitted. 

These discrepancies are significant and can lead to issues with physical fidelity when put into practical use as part of the 10-moment model.
\subsection{Closure Discovery}\label{sec:closure_discovery}
\begin{figure*}
    \centering
    \begin{multicols}{2}
    	\includegraphics[width=\linewidth]{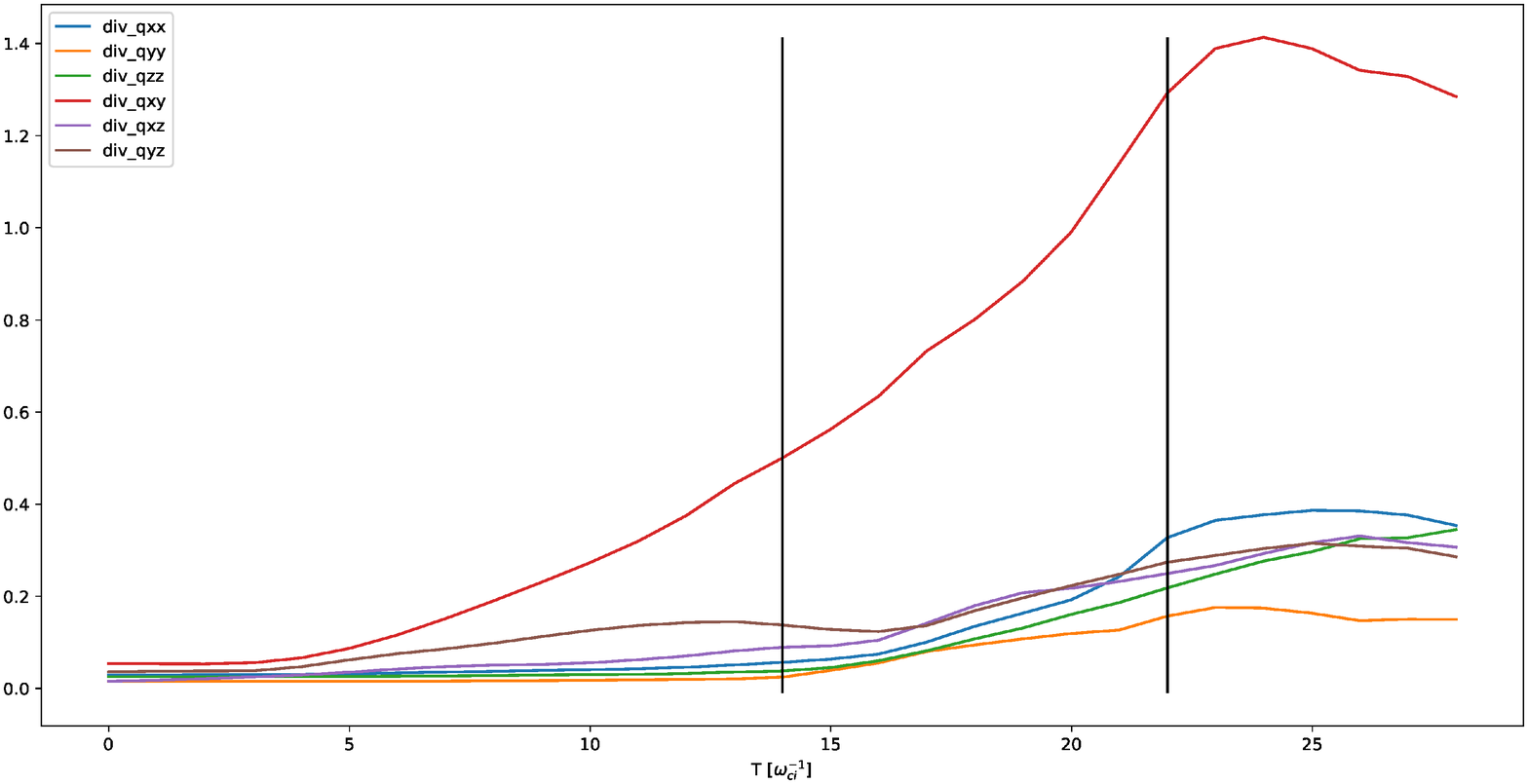}  \par
        \caption{The L-2 Norm of each component of the heat flux divergence tensor, taken across the entire simulation domain. Vertical bars represent the nonlinear phase of reconnection.}
        \label{fig:l2_div_qii}
    	\includegraphics[width=\linewidth]{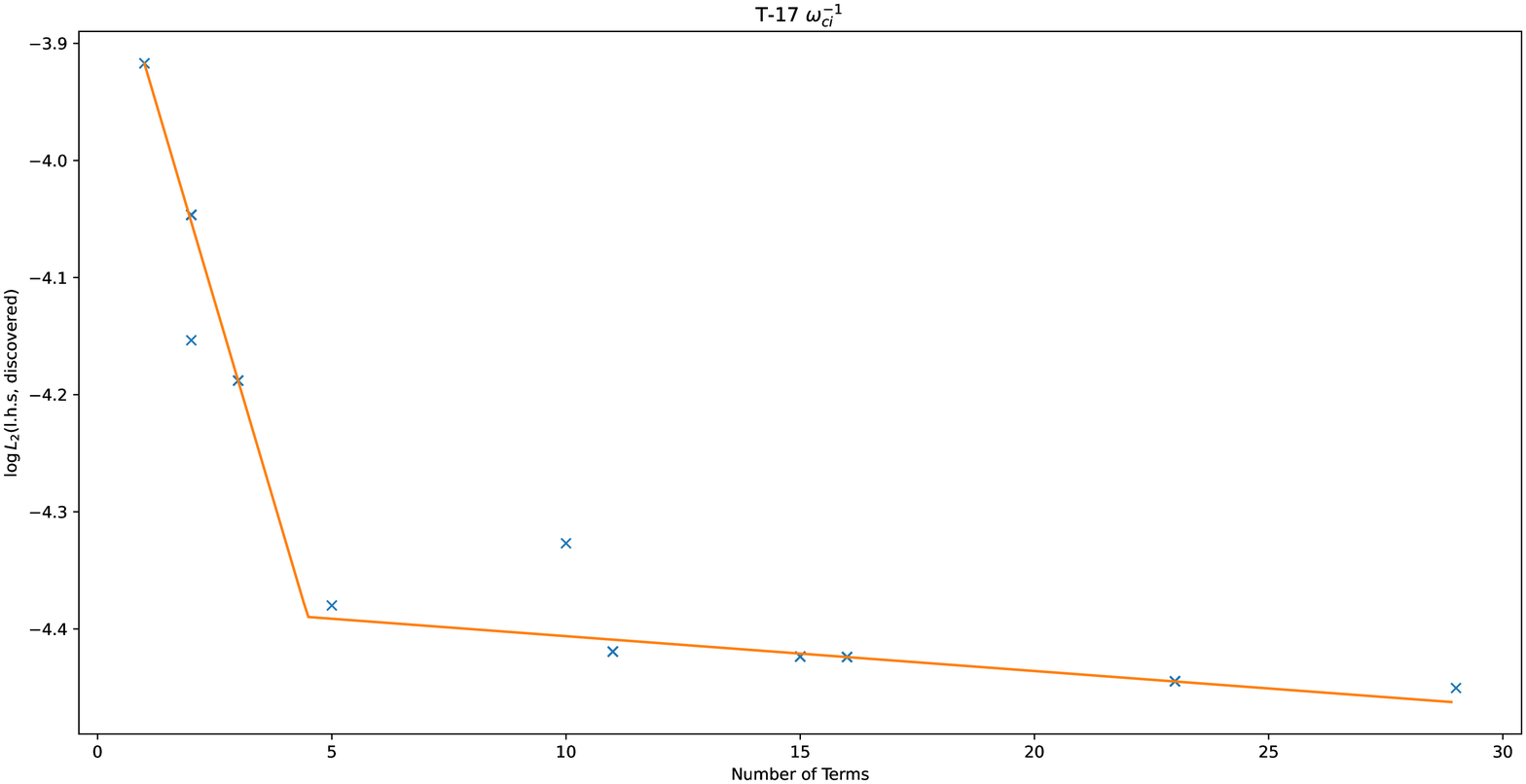} \par
    	\caption{A representative pareto curve constructed by solving SINDy with various upper and lower bounds for the yy-component of heat flux divergence.  The bounds that yield the elbow are selected for model discovery.}
    	\label{fig:pareto_curve}
   	\end{multicols}
\end{figure*}

Applying our methodology to the improvement or replacement of the above given closure, we decided to restrict ourselves to the nonlinear phase of the reconnection process.  This range may be seen in~\cref{fig:l2_div_qii}.
Pareto-optimal bounds are required for SINDy to discover a parsimonious solution.  Constructing a separate pareto curve for each element of $\nabla Q_{ij}$, we modified~\cref{alg:1} to that shown in~\cref{alg:2}.  A representative pareto curve used in discovery of the yy-component closure is shown in~\cref{fig:pareto_curve}.
\begin{algorithm}
\caption{Equation Synthesis}
\begin{algorithmic}[1]

\State $E, B, f^i \gets \text{Load field and moment data}$
\State Calculate $\partial_tf^i$, $\partial_xf^i$, $\partial_yf^i$, $\partial_zf^i$ for relevant moments
\State Integrate and construct term library $\Theta$
\For {\texttt{i, j \text{in} $\{x, y, z\} \times \{x, y, z\}$ } }
    \State calculate pareto-optimal bounds
    \State define $\nabla Q_{ij}$ as regression target
    \State Apply SINDy
\EndFor

\Return symbolic equation
\end{algorithmic}
\label{alg:2}
\end{algorithm}

In $\Theta$ we include each component of the pressure tensor scaled by the thermal velocity, the energy transfer term $\Vec{j}\cdot\Vec{E}$, the Poynting flux terms $\nabla \cdot \Vec{S}$, each component of the tensor $(\vec{u}\cdot\nabla)P$, and each term in the scalar product $(P\nabla)\cdot\vec{U}$.  These  terms were chosen for inclusion in $\Theta$ as each describes an energy flux and thus has the same physical units as heat flux divergence.

\begin{figure}
    \centering
    \includegraphics[width=\linewidth]{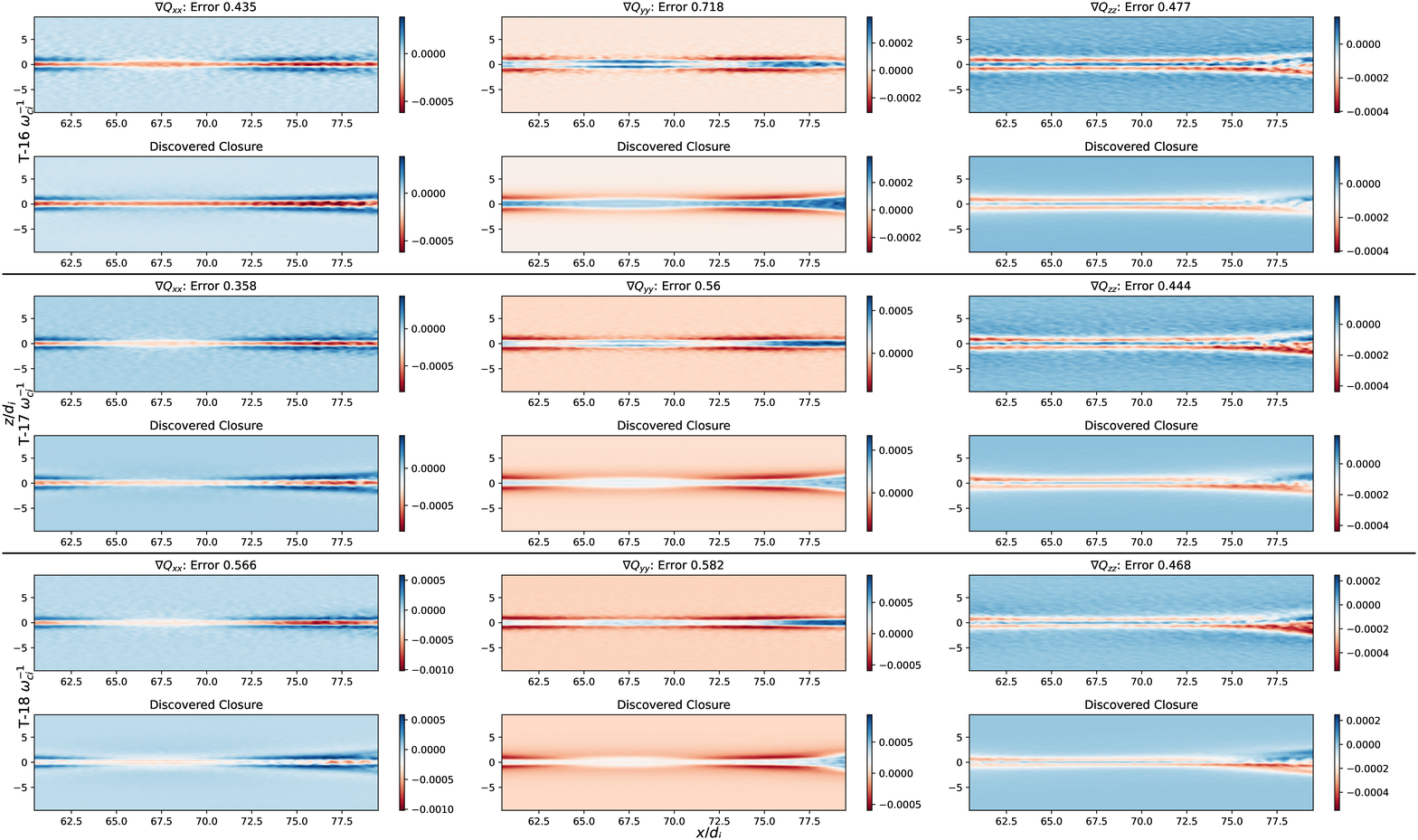}
    \caption{Snapshots of the discovered closure taken at intervals of 1 $w_{ci}^{-1}$ throughout the nonlinear phase, represented row-wise.  The diagonal components of $\nabla Q_{ii}$ are presented column-wise with the associated error in each of the headers.  Of note is the consistent performance of the closure as the nonlinear phase progresses.}
    \label{fig:closure_discovered}
\end{figure}

\Cref{tab:closureresults} summarizes our findings for a representative step (T-17 $\omega_{ci}^{-1}$) during the nonlinear phase and~\cref{fig:closure_discovered} demonstrates the temporal validity of the closure.  

The analysis was repeated at various times through the nonlinear phase and the discovered solutions did not vary greatly.  Thus, we present the predictions of the closure given by~\Cref{tab:closureresults} at evenly spaced time intervals through the nonlinear phase with the associated errors given in the headers of~\cref{fig:closure_discovered}.

The on-diagonal closures were able to reduce the error significantly when compared to the given closure.  The off-diagonal terms demonstrated no such improvement over the given Hammett-Perkins closure.  However, our proposed closure comes with some caveats to be discussed further in \S\ref{sec:discussion}.  

\begin{table}
\centering
\resizebox{\textwidth}{!}{\begin{tabular}{|c|c|c|c|c|}

    component & Hammett-Perkins & discovered & $L_2$($\nabla Q_{ij}$, HP) & $L_2$($\nabla Q_{ij}$, discovered)  \\ \hline
    xx & $0.02v_t(P_{xx} - p)$ & $2.85 k_0 v_t P_{yy} - 3.11 k_0 v_t P_{zz} + 1.70 P_{xx} \partial_x u_x - 1.33 j_y E_y + 2.52 u_z \partial_z P_{xx} + 3.28 u_x \partial_x P_{zz}$ & 0.995 & 0.358 \\ \hline
    
    yy & $0.04v_t(P_{yy} - p)$ & $0.80 k_0 v_t P_{zz} + 1.27 P_{xz} \partial_x u_x + 1.46 P_{xz} \partial_z u_z - 2.83 u_z \partial_z P_{xx} + 1.51 u_x \partial_x P_{zz}$  & 1.003 & 0.560 \\ \hline
    
    zz & $0.19v_t(P_{zz} - p)$ & $-1.09 k_0 v_t P_{xz} - 2.58 P_{yz} \partial_z u_z + 2.84 P_{xz} \partial_x u_x - 1.83 P_{xz} \partial_z u_z + 1.17 u_x \partial_x P_{xz} + 4.30 u_z \partial_z P_{xz} + 1.89 u_x \partial_x P_{yz} - 1.06 u_x \partial_x P_{zz} - 3.94 u_z \partial_z P_{zz}$ & 0.678 & 0.444\\ \hline
    
    xy & $2.60v_tP_{xy}$ & $1.32 u_z \partial_z P_{xy}$ & 1.120 & 1.000\\ \hline
    
    xz & $0.41v_tP_{xz}$ & $0.73 k_0 v_t P_{xz} - 3.59 u_z \partial_z P_{yz}$ & 1.023 & 0.997\\ \hline
    
    yz & $1.09v_tP_{yz}$ & $2.29 k_0 v_t P_{xz} - 3.63 P_{xz} \partial_x u_x - 2.57 u_x \partial_x P_{xz}$ & 0.941 & 0.993\\ \hline
\end{tabular}}
\caption{For each component of the heat flux divergence tensor we give the local approximate Hammett-Perkins closure, the discovered closure, and the relevant $L_2$ errors.  The $k_0$ for each component of Hammett-Perkins is calculated by averaging $\frac{\nabla Q_{ij}}{v_t(P_{ij} - p\delta_{ij})}$ across the domain.}
\label{tab:closureresults}
\end{table}

\section{Discussion and Conclusion}\label{sec:discussion}
Fully kinetic data may be used to verify the 10-moment multi-fluid model as demonstrated up to and including 2nd order conservation laws.  Specifically, 0th, 1st, and 2nd order equations were satisfied under visual inspection and with numerical error $L_2$(l.h.s, r.h.s)$\leq 0.28$ implying the multi-fluid 10-moment model may be validated with kinetic data.  This is evidence that the SINDy methodology may be applied to kinetic data, supporting the findings of~\citet{Alves2020}.

Applying the SINDy methodology, we were able to validate SINDy as a candidate data-driven approach up to and including the on-diagonal 2nd order terms as evidenced by the low $L_2$ coefficient error of the discovered solutions.  Difficulties were encountered in the discovery of the off-diagonal 2nd order r.h.s terms (as demonstrated by high coefficient error), where the discovered solutions narrowly deviated from the conservation laws but found more parsimonious solutions with lower numerical error.  This is due to the excluded terms having a low order of magnitude leading to elimination by SINDy and not evidence of violation of the conservation laws.   

Applied to heat flux divergence, this data-driven approach demonstrates strong results.  The existing local approximate Hammett-Perkins closure gives us a baseline to measure improvement.  The on-diagonal terms of the heat flux divergence were modeled with significantly reduced numerical $L_2$ error over the baseline.  Furthermore the discovered closures appear to hold through a significant portion of the nonlinear phase as demonstrated in~\cref{fig:closure_discovered}.  The off-diagonal terms were unsuccessfully modeled and further work is needed to understand this discrepancy.

%The off-diagonal terms were unsuccessfully modeled.  We believe this can be attributed to the high data noise, which is demonstrated in~\cref{tab:2ndresults}.  The off-diagonal 2nd order equations were predicted with an order of magnitude higher error than their on-diagonal counterparts.  If the data is simply too noisy to accurately represent the 2nd order off-diagonal terms, we cannot expect the off-diagonal closure modeling to succeed. 

Other shortcomings of this approach lie in perturbation phase restrictions.   While the discovered closure holds well through the the nonlinear phase, there is no such guarantees during the linear phase.  The local approximate Hammett-Perkins closure was derived using quasi-linear theory, as such we conjecture that the discovered closure will apply in the linear regime as well.  Unfortunately we cannot demonstrate this at this time, due to the low signal-to-noise ratio in the linear phase. %Further work needs to be done in either incorporating the linear phase into the regression, finding a similar set of closures which may be applied during the linear phase.

As a data-driven method we make no attempt to explain the physics of the discovered closures here.  Our sole claim is that the discovered closures will yield improved physical fidelity during the nonlinear phase when compared to the local Hammett-Perkins closure.  Model validation is paramount when working with data driven methodologies, as such future directions in this research will involve applying the discovered closure to a multi-fluid simulation where we hope to see improvements over local Hammett-Perkins closure.  Further, we will be looking at expanding the methodology with the goal of pursing a closure which accurately represents the off-diagonal terms.  

\subsection*{Declaration of Interests}
The authors report no conflict of interest. 
\subsection*{Funding}
This research received no specific grant from any funding agency, commercial or not-for-profit sectors.

\bibliographystyle{jpp}

\bibliography{jpp-submission}

\end{document}